\documentclass[10pt,a4paper,fleqn]{article}
\usepackage[T1]{fontenc}
\usepackage{jheppub}
\usepackage{alemfonts}
\usepackage[utf8]{inputenc}
\usepackage[english]{babel}
\usepackage{amsmath}
\usepackage{amsfonts}
\usepackage{amssymb}
\usepackage{xcolor}
\usepackage{mathrsfs}
\usepackage{amsthm}
\usepackage{float}
\usepackage[parfill]{parskip}

\long\def \beq#1\eeq {\begin{equation} #1 \end{equation}}
\long\def \beaq#1\eeaq {\begin{equation}\begin{aligned} #1 \end{aligned}\end{equation}}
\long\def \bes#1\ees {\begin{equation}\begin{split} #1 \end{split} \end{equation}}
\long\def \bea#1\eea {\begin{eqnarray} #1 \end{eqnarray}}
\long\def \bse[#1]#2\ese {\begin{subequations}\label{#1}\begin{align} #2 \end{align}\end{subequations}}
\newcommand{\der}[2]{\frac{\partial #1}{\partial #2}}
\newcommand{\evalat}[2]{\left. #1 \right\vert_{#2}}

\newcommand{\mv}[1]{\langle #1\rangle}

\newcommand{\qE}[1]{\operatorname{\mathbb{E}_{#1}}}
\long\def\dm[#1]{\!\operatorname{d\mu}\left(#1\right)}

\newcommand{\RS}{{\rm \scriptscriptstyle RS}}
\theoremstyle{plain}
\newtheorem{Remark}{Remark}
\newtheorem{Theorem}{Theorem}

\newtheorem{Proposition}{Proposition}

\newtheorem{Definition}{Definition}

\graphicspath{{./Figures/}}

\title{Dreaming neural networks: rigorous results}
\author[a,b]{Elena Agliari}
\author[c]{Francesco Alemanno}
\author[c,d,e]{Adriano Barra}
\author[c,d,e]{Alberto Fachechi}

\affiliation[a]{Dipartimento di Matematica, Sapienza Universit\`a di Roma, Italy}
\affiliation[b]{GNFM-INdAM Sezione di Roma, Italy}
\affiliation[c]{Dipartimento di Matematica e Fisica Ennio De Giorgi, Universit\`a del Salento, Italy}
\affiliation[d]{GNFM-INdAM Sezione di Lecce, Italy}
\affiliation[e]{INFN, Istituto Nazionale di Fisica Nucleare, Sezione di Lecce, Italy}

\emailAdd{alberto.fachechi@le.infn.it}
\emailAdd{elena.agliari@uniroma1.it}
\emailAdd{adriano.barra@unisalento.it}
\emailAdd{francesco.alemanno@le.infn.it}
\abstract{Recently a {\em daily routine} for associative neural networks has been proposed: the network Hebbian-learns during the {\em awake state} (thus behaving as a standard Hopfield model), then, during its {\em sleep state}, optimizing information storage, it consolidates pure patterns and removes spurious ones: this forces the synaptic matrix to collapse to the projector one (ultimately approaching the Kanter-Sompolinksy model). This procedure keeps the learning Hebbian-based (a biological must) but, by taking advantage of a (properly stylized) sleep phase, still reaches the maximal critical capacity (for symmetric interactions).
\newline
So far this emerging picture (as well as the bulk of papers on {\em unlearning techniques}) was supported solely by mathematically-challenging routes, e.g. mainly replica-trick analysis and numerical simulations: here we rely extensively on Guerra's interpolation techniques developed for neural networks and, in particular, we extend the generalized stochastic stability approach to the case. Confining our description within the replica symmetric approximation (where the previous ones lie), the picture painted regarding this generalization (and the previously existing variations on theme) is here entirely confirmed.
\newline
Further, still relying on Guerra's schemes, we develop a systematic fluctuation analysis to check where ergodicity is broken (an analysis entirely absent in previous investigations). Remarkably we find that, as long as the network is awake, ergodicity is bounded by the Amit-Gutfreund-Sompolinsky critical line (as it should), but, as the network sleeps, sleeping destroys spin glass states by extending both the retrieval as well as the ergodic region: after an entire sleeping session the solely surviving regions are retrieval and ergodic ones and this allows the network to achieve the {\em perfect retrieval regime} (where the number of storable patterns exactly equals the number of neurons the network is built of).
}
\begin{document}
\maketitle
\section{Introduction}

\begin{figure}[!]
	\centering
	\begin{minipage}[c]{.7\textwidth}
		\centering
		\includegraphics[width=\textwidth]{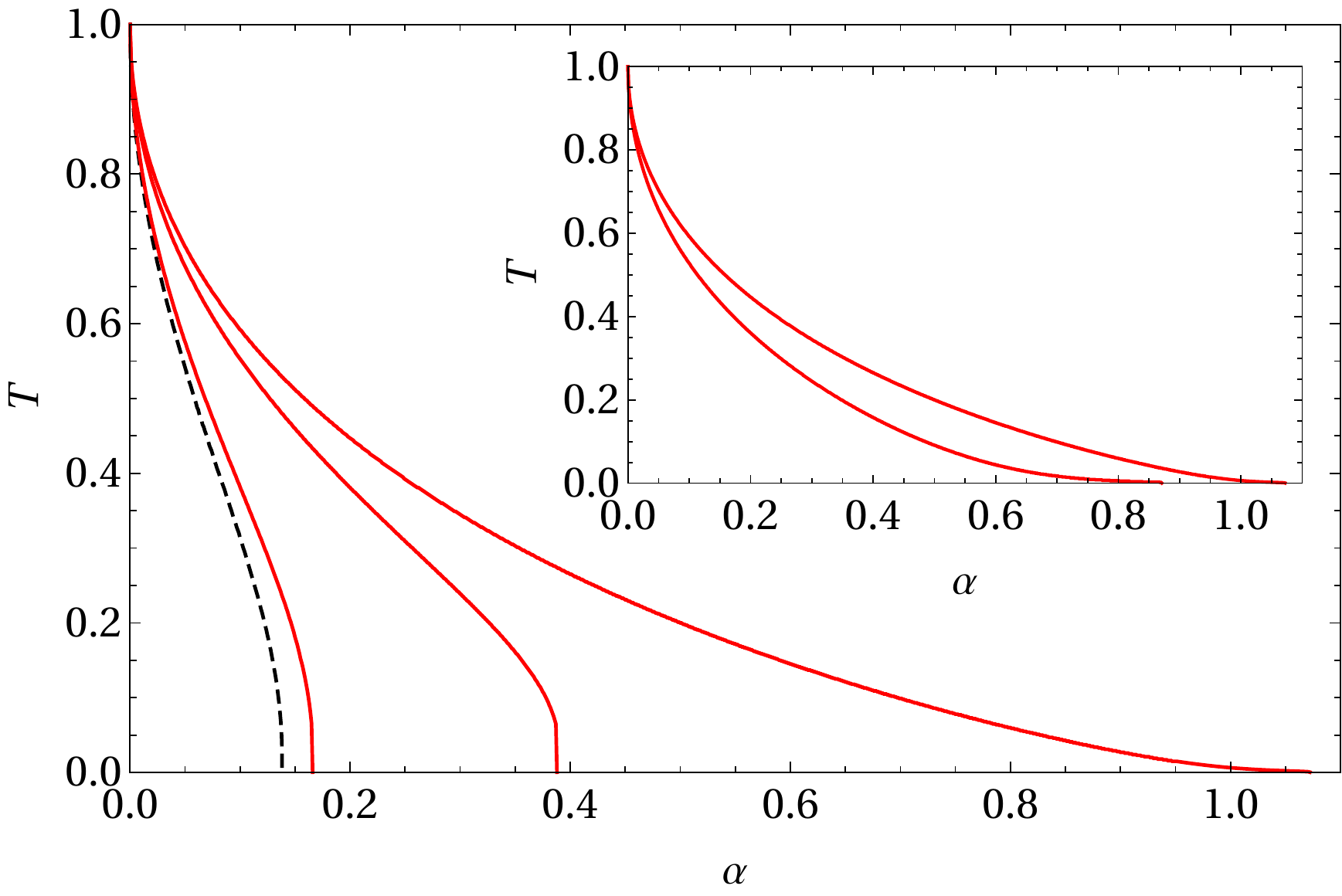}
	\end{minipage}%
	\caption{Critical line for the transition between retrieval and spin-glass phases for various values of the unlearning time. From the left to the right: $t=0$ (Hopfield, black dashed line), $0.1$, $1$ and $1000$. The inset shows two curves tracing the boundary of the maximal retrieval regions where patterns are global free energy minima (inner boundary) or local free energy minima (outer boundary) in the long sleep limit.}\label{fig:criticallines}
\end{figure}

\begin{figure}[H]
	\centering
	\includegraphics[width=0.7\textwidth]{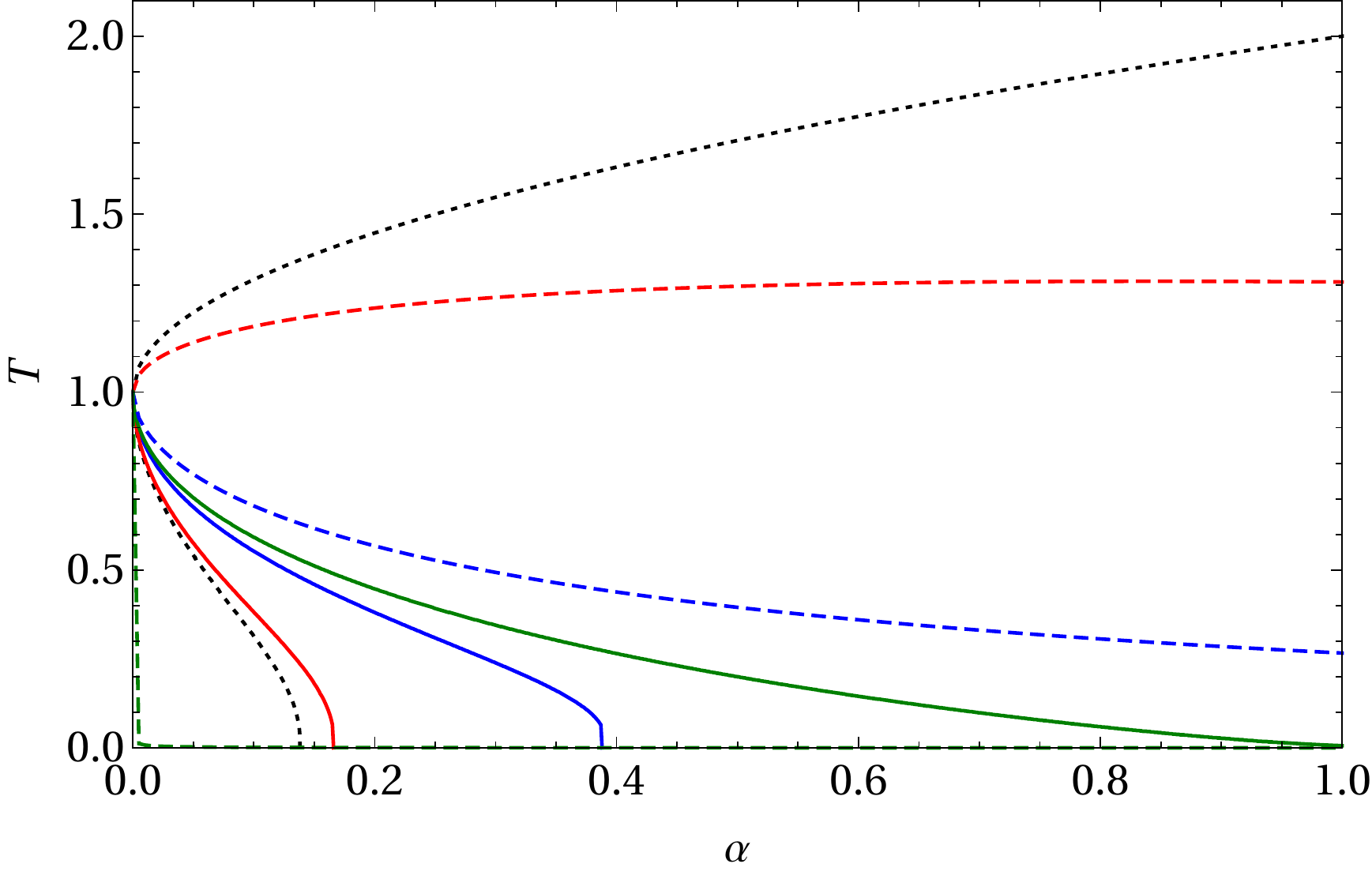}
	\caption{Critical lines for ergodicity breaking (dotted curves) and retrieval region boundary (solid curves) for various values of the unlearning time. From the top to the bottom: $t=0$ (black lines, i.e. the Hopfield phase diagram), $t=0.1$ (red lines), $1$ (blue lines) and $1000$ (green lines).}\label{fig:erglines}
\end{figure}

\begin{figure}[H]
	\centering
	\begin{minipage}[c]{1.0\textwidth}
		\centering
		\includegraphics[width=\textwidth]{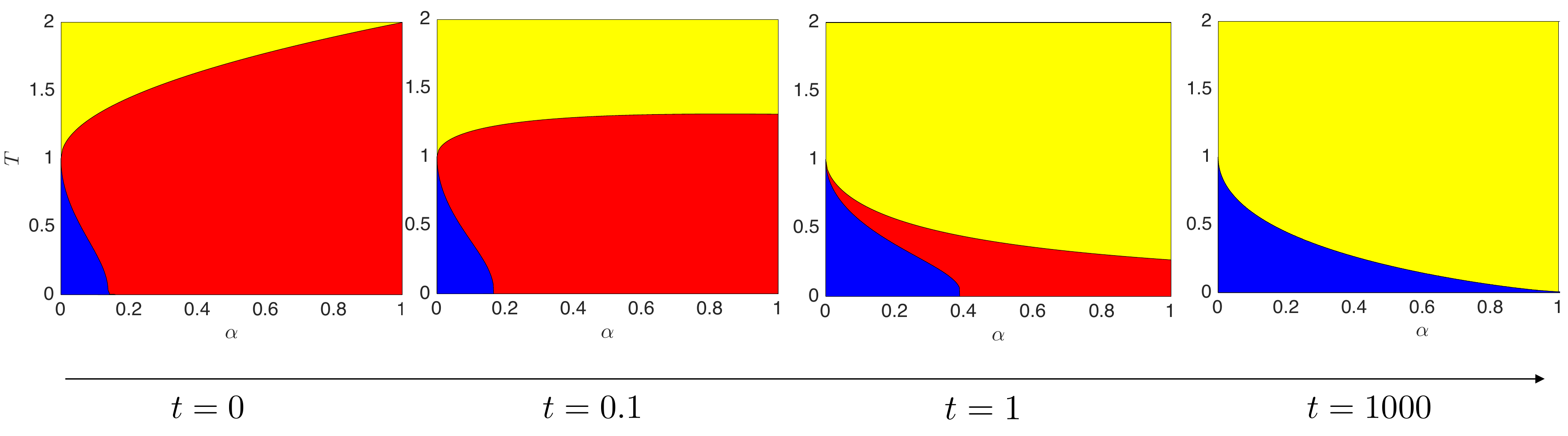}
	\end{minipage}%
	\caption{The phase diagram is depicted for different choices of $t$, namely, from left to right, $t=0, 0.1, 1, 1000$. Notice that, as $t$ grows, the retrieval region (blue) and the ergodic region (yellow) get wider at the cost of the spin-glass region (red) which progressively shrinks up to collapse as $t \rightarrow \infty$. Also notice the change in the concavity of the critical line which separates ergodic and spin-glass region.}\label{fig:phasediag}
\end{figure}

Statistical mechanics of spin glasses \cite{MPV} has been playing a primary role in the investigation of neural networks, as for the description of both their learning phase \cite{angel-learning,sompo-learning} and their retrieval properties \cite{Amit,Coolen}. Along the past decades, beyond the bulk of results achieved via the so-called replica-trick \cite{MPV}, a considerable amount of rigorous results exploiting alternative routes (possibly mathematically more transparent) were also developed (see e.g. \cite{Agliari-Barattolo,ABT,Bovier1,Bovier2,Bovier3,Albert1,Barra-JSP2010,bipartiti,Dotsenko1,Dotsenko2,Tala1,Tala2,Tirozzi,Pastur}  and references therein). This paper goes in the latter direction and focuses on a generalization of the Hopfield model \cite{Albert2} that is able to saturate the optimal storage capacity and whose main characteristics are summarized hereafter.
\newline
In \cite{Albert2} the Hebbian kernel underlying the Hopfield model was revised to account also for {\it reinforcement} and {\it removal} processes. The resulting kernel can be interpreted as the effect of a {\it daily routine}: during the {\em awake} state, the network is fed with inputs (i.e. {\em patterns} of information) that are stored in an Hebbian fashion\footnote{We stress that, given the equivalence between restricted Boltzmann machines and Hopfield neural networks \cite{BarraEquivalenceRBMeAHN}, also learning via e.g. {\em contrastive divergence} \cite{Hinton1} ultimately falls into the Hebbian category \cite{Agliari-Dantoni,Agliari-Isopi}.}, then, during the {\em asleep} state, it weeds out the (combinatorial\footnote{The growth in the number of spurious states is roughly exponential in the number of stored patterns, namely -in the high storage regime- in the number of neurons.}) proliferation of the spurious mixtures (unavoidably created as metastable states in the free-energy landscape of the system during the learning stage) and it consolidates the pure states (making their free-energy minima deeper in this landscape picture). Remarkably, after these procedures, the network is able to saturate the storage capacity $\alpha$ (that is the amount of stored patterns $P$ over the amount of available neurons $N$, in the thermodynamic limit, i.e. $\alpha = \lim_{N \to \infty}P/N$) to its upper bound\footnote{Actually the network seems to perform even {\em better}, returning its maximal capacity to be $\alpha_c \sim 1.07 > 1$: this is obviously not possible and, as explained by Dotsenko and Tirozzi \cite{Dotsenko1,Dotsenko2}, it is a chimera of the replica-symmetric regime at which the theory is developed.} which, for symmetric networks, is $\alpha_c=1$ \cite{Gardner}. Further, in the retrieval phase of its parameter space, pure states are global minima up to $\alpha \sim 0.85$ (see Figure \ref{fig:criticallines}), that is a much broader range with respect to the classical Hopfield counterpart, where they remain global minima solely for $\alpha < 0.05$.
\newline
In this work, we first show the equivalence between the aforementioned generalized neural network and a tripartite (or ``three-layers'' in a machine-learning jargon) spin-glass, where couplings between neurons of different layers exhibit correlations and the third layer is a {\em spectral layer} equipped with imaginary numbers (see Fig. \ref{fig:GeneralizedBoltmannMachine} and Remark \ref{remark:quarto}).
Then, we generalize the stochastic stability technique, introduced in \cite{AC1,CG1} to address Sherrington-Kirkpatrick spin-glass and later developed in \cite{Barra-JSP2010} to account also for bipartite spin-glasses (namely restricted Boltzmann machines or Hopfield networks \cite{BarraEquivalenceRBMeAHN} in a machine learning jargon \cite{DLbook,Hugo}), so that it can as well deal with the present tripartite and correlated spin-glass.
\newline
Next, by using this novel approach -that is mathematically well controllable at any stage of the calculations- we obtain the expression of the quenched replica-symmetric free energy related to the model  (as well as the set of self-consistent equations for the order parameters) and we show that the resulting picture sharply coincides with that obtained via the replica-trick analysis \cite{Albert2}. This implies, in a cascade fashion, that all the results previously heuristically derived are actually proved (the most remarkable one being the saturation of the critical capacity).
\newline
Finally, we extend our analysis to order-parameter fluctuations in order to investigate ergodicity breaking: interestingly, as suggested also by the self-consistencies, we find that -without sleeping- ergodicity breaks as predicted by Amit-Gutfreund-Sompolinsky \cite{Amit} (as it should), but -as sleeping takes place- the spin-glass region shrinks and ultimately the network phase-diagram exhibits only retrieval and ergodic phases  (see Fig.s \ref{fig:erglines},\ref{fig:phasediag}).
%

This paper is structured as follows: in Sec.~$2$, once the model is introduced and embedded in its statistical mechanical framework, we calculate  its quenched free energy by introducing a novel interpolating structure \`a la Guerra and this provides a first picture of the phase diagram of the model (as we can identify the transition between the retrieval and the spin-glass regions). Next, in Sec.~$3$, we study the fluctuations of the order parameters to inspect where ergodicity is spontaneously broken as this is a signature of the critical line, namely the transition between the ergodic and the spin-glass regions): by combining the two results a full picture of the phase diagram of the model can be finally deduced. Sec.~$4$ is left for conclusions. Technical details and further remarks on the interpolation approach are provided in the appendices.

\section{Replica symmetric free energy analysis}
\subsection{Definition of the Model}\label{replica-simmetric-theory}
Driven by the works of Personnaz, Guyon, Dreyfus \cite{Personnaz} and of Dotsenko et al. \cite{Dotsenko1,Dotsenko2}, in \cite{Albert2} we introduced the following generalization of the standard Hopfield paradigma \cite{Hopfield}, referred to as ``reinforcement\&removal'' (RR) algorithm:
consider a network composed by $N$ Ising neurons $\{ \sigma_i \}_{i=1,...,N}$ and $P$ patterns $\{\xi^{\mu}\}_{\mu=1,...,P}$ (namely quenched random vectors of the same length $N$), and denote with $t \in \mathbb{R}^+$ the sleep extent (such that for $t=0$ the network has never slept, while for $t\to \infty$ an entire sleeping session has occurred), we can then introduce the following
\begin{Definition}
The  Hamiltonian of the reinforcement$\&$removal model reads as:\footnote{As a matter of notation, we stress that the denominator $1/(\mathbb I+tC)$ in the generalized kernel is intended as the inverse matrix $(\mathbb I+tC)^{-1}$.}
\beq\label{new-model}
H_{N,P}^{\textrm{(RR)}}(\sigma|\xi,t):= - \frac{1}{2N}\sum_{i=1}^{N}\sum_{j=1}^{N}\sum_{\mu=1}^{P}\sum_{\nu=1}^{P}\xi_i^{\mu}\xi_j^{\nu}\left( \frac{1+t}{\mathbb{I}+t C} \right)_{\mu,\nu} \sigma_i \sigma_j,
\eeq
where $\sigma_i = \pm 1 \  \forall i \in (1,...,N)$, $\xi^1$  -that is the pattern candidate to be retrieved- has binary entries $\xi_i^{1} \in \{-1,+1\}$ drawn from $P(\xi_i^{\mu}=+1) = P(\xi_i^{\mu}=-1) = \frac12$, while the remaining $P-1$ patterns $\{\xi^{\mu}\}_{\mu=2,...,P}$, have i.i.d. standard Gaussian entries $\xi_i^{\mu} \sim \mathcal{N}[0,1]$, and the correlation matrix $\boldsymbol{C}$ is defined as
$$
C_{\mu,\nu} := \frac{1}{N}\sum_{i=1}^{N}\xi_i^{\mu}\xi_i^{\nu}.
$$
\end{Definition}

\begin{Remark}\label{Marco2}
We stress that, for the sake of mathematical convenience, as deepened in \cite{Agliari-Barattolo}, we take solely the pattern candidate for retrieval (i.e. the {\em signal}) to be Boolean, while all the remaining ones (acting as {\em slow noise} on the retrieval) are chosen as Gaussian: although neural networks, in general, do not exhibit the universality properties of spin glasses \cite{Genovese}, this is no longer true if we confine our focus solely to the structure of the slow noise generated by patterns\footnote{As extensively discussed in \cite{Barra-RBMsPriors1,Barra-RBMsPriors2} by varying the nature of the neurons as well as of the pattern entries, for instance ranging from Boolean (Ising) to standard Gaussians, the retrieval performances of the network vary sensibly and, in some limits, are entirely lost: in this sense neural networks do not share {\em universality} with standard spin-glasses.}. 
\end{Remark}

\begin{Remark}
Note that the matrix ${\xi}^T \left( \frac{1+t}{\mathbb{I}+ t {C}} \right) {\xi}$, encoding the neuronal coupling, recovers the Hebbian kernel for $t=0$ , while it approaches the pseudo-inverse matrix for $t \rightarrow \infty$  (see \cite{Albert2} for the proof). Accordingly, the model described by the Hamiltonian (\ref{new-model}) spans, respectively, from the standard Hopfield model  $(t \to 0)$ to the Kanter-Sompolinksy model \cite{KanterSompo}  $(t \to \infty)$.
\newline
During the sleeping session, both reinforcement and remotion take place: oversimplifying, in the generalized synaptic coupling appearing in (\ref{new-model}), the denominator ({\it i.e.}, the term $\propto (1+tC)^{-1}$) yields to the remotion of unwanted mixture states, while the numerator ({\it i.e.}, the term $\propto 1+t$) reinforces the pure memories.
\end{Remark}

We are interested in obtaining the phase diagram of the model coded by the cost function (\ref{new-model}), solely in the thermodynamic limit and under the replica symmetric assumption. To achieve this goal the following definitions are in order.
\begin{Definition}
Using $\beta \in \mathbb{R}^+$ as a parameter tuning the level of {\em fast noise} in the network (with the physical meaning of inverse temperature, i.e. calling $T$ the temperature, $\beta \equiv T^{-1}$ in proper units,), the partition function of the  model (\ref{new-model}) is introduced as
\beq\label{lazoccola}
Z_{N,P}(\sigma|\xi,t)  := \sum_{\{\sigma\}} e^{-\beta H_{N,P}^{(RR)}(\sigma|\xi,t)} = \sum_{\{ \sigma \}}\exp\left[ \frac{\beta}{2N }\sum_{i,j=1}^{N,N}\sum_{\mu,\nu=1}^{P,P}\xi^\mu _i \xi ^\nu _j \left( \frac{1+t}{\mathbb{I} +t  C}\right)_{\mu,\nu} \sigma_i \sigma_j\right].
\eeq
\end{Definition}
\begin{Definition}
Denoting with $\mathbb{E}_{\xi}$ the average over the quenched patterns, for a generic function $O(\sigma,\xi)$ of the neurons and the couplings, we can define the Boltzmann $\langle O(\sigma,\xi)\rangle$ as
\begin{eqnarray}
\langle O(\sigma,\xi)  \rangle &:=& \frac{\sum_{\{\sigma\}} O(\sigma,\xi) e^{-\beta H^{(RR)}_{N,P}(\sigma|\xi,t)}}{Z_{N,P}(\sigma|\xi,t)},\\
\end{eqnarray}
such that its quenched average reads as $\mathbb{E}_{\xi} \langle O(\sigma,\xi) \rangle$.
\end{Definition}
\begin{Definition}
Once introduced the partition function $Z_{N,P}(\sigma|\xi,t)$, we can define the infinite volume limit of the intensive quenched  free-energy $F_N(\alpha, \beta, t)$ and of the intensive quenched pressure $A(\alpha,\beta,t)$ associated to the model (\ref{new-model}) as
\beq\label{freeEnergy}
- \beta F(\alpha, \beta, t) \equiv  A(\alpha,\beta,t) := \lim_{N \to \infty} \frac1{N} \mathbb{E} \ln  Z_{N,P}(\sigma|\xi,t).
\eeq
\end{Definition}

As anticipated, the pressure of the model (\ref{new-model}) was analyzed in \cite{Albert2} via replica-trick \cite{Coolen} (corroborated by extensive numerical simulations), showing that (at the replica symmetric level of description) the maximal critical capacity of this neural network saturates the Gardner's bound \cite{Gardner} (i.e. $\alpha_c =1$, for symmetric noiseless networks).

\begin{figure}[!]
	\centering
	\begin{minipage}[c]{.7\textwidth}
		\centering
		\includegraphics[width=\textwidth]{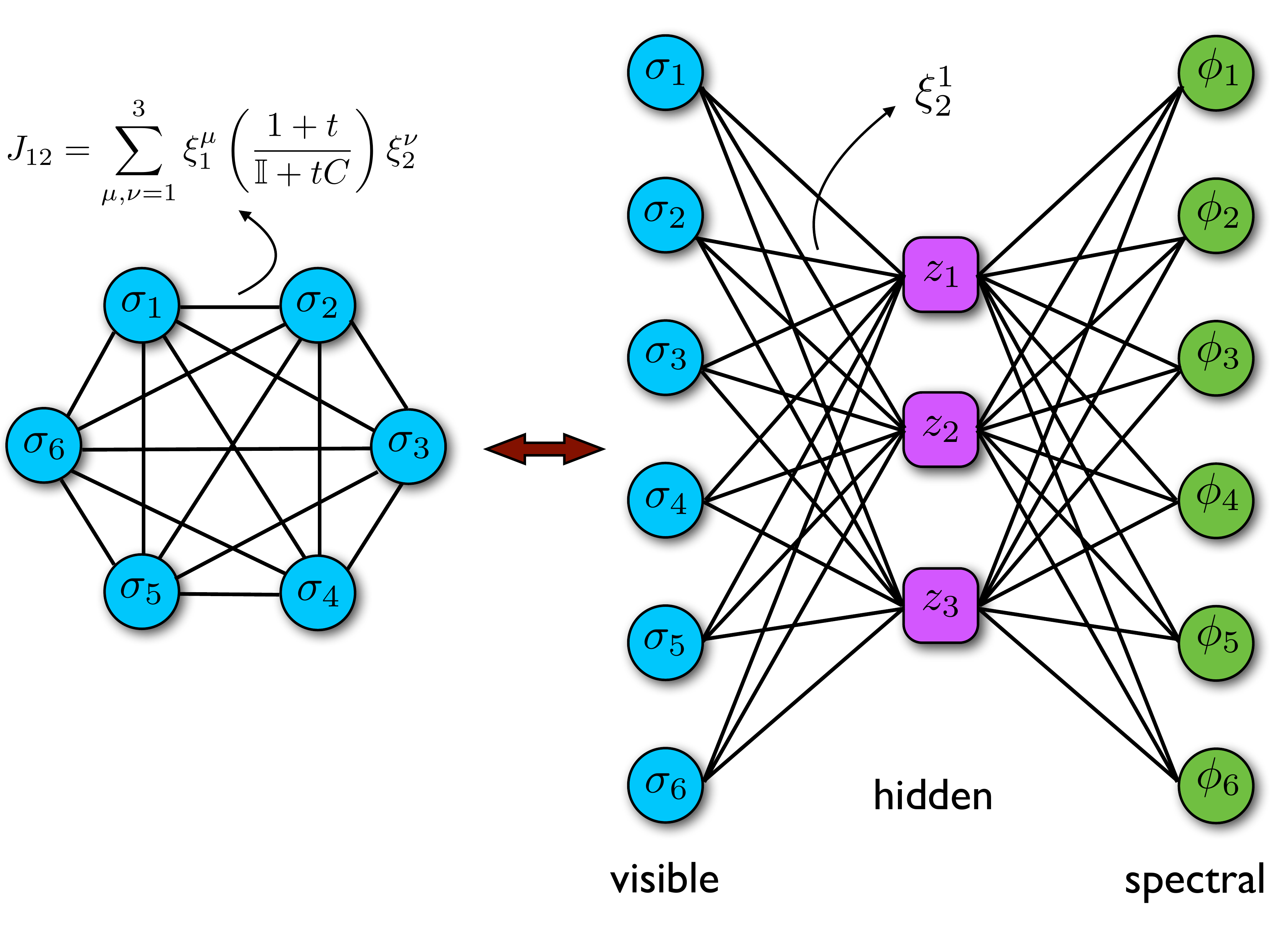}
	\end{minipage}%
	\caption{Stylized representation of the generalized Hopfield network (left) and its dual generalized (restricted) Boltzmann machine (right), namely the three-partite spin-glass under study: in machine learning jargon these parties are called {\em layers} and, here, they are respectively the visible, hidden and spectral layers. Note further that, as it should, when $t \to 0$ the  duality above reduces to the standard picture of Hopfield networks and restricted Boltzmann machines \cite{Agliari-Barattolo,BarraEquivalenceRBMeAHN}.}\label{fig:GeneralizedBoltmannMachine}
\end{figure}

\begin{Remark}\label{remark:quarto}
The partition function defined in (\ref{lazoccola}) can be represented in Gaussian integral form as
\beq\label{eq:boltzmann}
\begin{split}
 Z_{N,P}(\sigma|\xi,t) &=  \sum_{\{ \sigma \} } \int \Big(\prod_{\mu=1}^{P} d \mu(z_\mu) \Big)\Big(\prod_{i=1}^{N} d \mu(\phi_i) \Big)\cdot \\ &\cdot \exp\left(\sqrt{\frac{\beta}{N} (t+1)}\sum_{\mu, i}^{P,N}z_\mu  \xi^\mu _i \sigma_i  +i\sqrt{\frac{t}{N}} \sum_{\mu, i}^{P,N}z_\mu \xi^\mu _i \phi _i \right),
\end{split}
\eeq
where $d \mu(z_\mu)$ and $d \mu(\phi_i)$ are the standard Gaussian measures.
This relation shows that the partition function of the reinforcement$\&$removal model is equivalent to the partition function of a tripartite spin-glass where the intermediate party (or {\em hidden layer} to keep a machine learning jargon) is made of real neurons $\{ z_{\mu}\}_{\mu=1,...,P}$ with $z_{\mu} \sim \mathcal{N}[0,1],  \forall \mu$, while the external layers are made, respectively, of a set of Boolean neurons $\{\sigma_i \}_{i=1,...,N}$ (the {\em visible layer}) and of a set of imaginary neurons with magnitude $\{ \phi \}_{i=1,...,N}$, being $\phi_{i} \sim \mathcal{N}[0,1],  \forall i$ (the {\em spectral layer}), see Fig. \ref{fig:GeneralizedBoltmannMachine}.
\end{Remark}
\subsection{Guerra's interpolating framework for the free energy}
\begin{Definition}
Once expressed the partition function (\ref{lazoccola}) in its integral representation (\ref{eq:boltzmann}), we can introduce the related tripartite spin glass Hamiltonian as
\begin{equation}
H_{N,P}=\frac{a}{\sqrt{N}}\sum_{i=1}^N \sum_{\mu=1}^P  z_\mu \xi_i^\mu k_i,
\label{eq:energycostf}
\end{equation}
where we introduced the ``multi-spin'' $k_i=\sigma_i+b \phi_i$ and where
\beq
a=\sqrt{\beta(t+1)},\quad  b=i\sqrt{\frac{t}{\beta(t+1)}}.
\label{eq:defspins}
\eeq
\end{Definition}
\begin{Remark}
Note that the cost function (\ref{eq:energycostf}) and the one associated to the original model (\ref{new-model}) share the same partition function and therefore exhibit the same Thermodynamics. By a practical perspective, the latter is more suitable for understanding the retrieval capabilities of the network, the former for dealing with its learning skills \cite{BarraEquivalenceRBMeAHN,Barra-RBMsPriors1}.
\end{Remark}

In the following we consider the challenging case with $P=\alpha N$ for large $N$ and we aim to obtain an expression for the quenched pressure (\ref{freeEnergy}) in terms of the order parameters introduced in the next
\begin{Definition}
The natural order parameters for the neural network model (\ref{new-model}) -as suggested by its integral representation (\ref{eq:energycostf})- are the overlaps $q_{ab}$ and $p_{ab}$ between the $k$'s and the $z$'s variables, respectively, as functions of two replicas (a,b) of the system, and the generalized Mattis overlap\footnote{We arbitrarily (but with no loss of generality) nominated the first pattern as the retrieved one.} $m_1$, namely
\bes
q_{ab}&:=\frac{1}{N}\sum_{i=1}^N k_i^{(a)} k_i^{(b)},\\
p_{ab}&:=\frac{1}{P}\sum_{\mu\ge 2} z_\mu^{(a)} z_\mu^{(b)},\\
m_1&:=\frac{1}{N}\sum_{i=1}^N \xi_i^1 k_i.
\label{eq:orderparam}
\ees
\end{Definition}
\begin{Remark}
The replica symmetric approximation (RS) is imposed by requiring that the order-parameters of the theory do not fluctuate in the thermodynamic limit\footnote{This request is obviously perfectly consistent with the replica-symmetric ansatz when approaching the problem via the replica trick \cite{Coolen,Albert2}.}, i.e.
\bes
q_{ab}& \overset{\rm \scriptscriptstyle RS}{\to} W \delta_{ab}+ q (1-\delta_{ab}),\\
p_{ab}& \overset{\rm \scriptscriptstyle RS}{\to} X \delta_{ab}+ p (1-\delta_{ab}),\\
m_1& \overset{\rm \scriptscriptstyle RS}{\to} m,
\label{eq:rsorderparam}
\ees
where we called, respectively, $W,q,X,p,m$ the replica symmetric values of the diagonal and off-diagonal overlap $q$, the diagonal and off-diagonal overlap $p$ and the Mattis magnetization $m_1$.
\end{Remark}

\bigskip

Now the plan is to get an explicit expression for the pressure (\ref{freeEnergy}) in terms of these order parameters, to extremize the former over the latter and get a phase diagram for the network. To reach this goal we generalize a Guerra's interpolation scheme \cite{Barra-JSP2010}: the idea is to compare the original system, as represented in eq. (\ref{eq:energycostf}) (namely a three-layer correlated spin glass), with three random single-layers, where each layer experiences, statistically, the same mean-field that would have been produced by the other layers over it. To this aim we introduce the following  
\begin{Definition}
Being $s \in [0,1]$ an interpolating parameter, $\{ \eta_i \}_{i \in (1,...,N)}$ a set of $N$ i.i.d. Gaussian variables, $\{ \lambda_{\mu} \}_{\mu \in (2,...,P)}$ a set of $P-1$ i.i.d. Gaussian variables, and the scalars $C_1,C_2,C_3,C_4,C_5$ to be set a posteriori, we use as interpolating pressure the following quantity
\begin{eqnarray}
\label{eq:interpfunc}
\mathcal{A}(s)&:=&\frac{1}{N}\qE{\xi,\eta,\lambda} \ln\sum_{\sigma}\int\dm[z,\phi] \exp \Big[ \sqrt{s}\frac{a}{\sqrt{N}}\sum_{i,\mu\ge 2} z_\mu \xi_i^\mu k_i+\sqrt{s}\frac{a}{\sqrt{N}}\sum_{i} z_1 \xi_i^1 k_i\\
& + &\sqrt{1-s}\Big( C_1 \sum_i^N \eta_i k_i + C_2 \sum_{\mu\ge2} \lambda_\mu z_\mu \Big)+\frac{1-s}{2}\Big( C_3 \sum_{\mu\ge 2} z_\mu^2 + C_4 \sum_i k_i^2 + C_5 a \sum_i \xi_i^1 k_i \Big)\Big].
\nonumber
\end{eqnarray}
\end{Definition}
\begin{Remark}
When $s=1$ we recover the original model, namely $A(\alpha,\beta,t)=\lim_{N \to \infty}\mathcal{A}(s=1)$, while for $s \to 0$ we are left with a one-body problem, and, consequently, the probabilistic structure of $\mathcal{A}(s=0)$ is more tractable.
\end{Remark}
\begin{Remark}
We note the importance of splitting the sum on the $\xi$'s into $\xi^1$ (i.e. the {\em signal}) and the $\xi^2\cdots\xi^P$ (i.e. the {\em quenched noise}) since the quenched average treats them differently, and so we will need to address them separately.
\end{Remark}
\begin{Proposition}
The infinite volume limit of the quenched pressure related to the model (\ref{new-model}) can be obtained by using the Fundamental Theorem of Calculus as
\begin{equation}
A(\alpha,\beta,t)\equiv \lim_{N \to \infty} \mathcal{A}(s=1)= \lim_{N \to \infty} \left( \mathcal{A}(s=0)+\int_0^1 \frac{d \mathcal{A}(s)}{ds}\, ds\right).
\label{eq:sumrule}
\end{equation}
\end{Proposition}
To follow this approach, two calculations are in order: the streaming $d_s \mathcal{A}(s)$ (and its successive back-integration) and the evaluation of the Cauchy condition $\mathcal{A}(s=0)$. Let us start with $d_s \mathcal{A}(s)$:
\bes
\frac{d \mathcal{A}(s)}{ds}=&\frac{1}{2N}\qE{\xi,\lambda,\eta}\Big[ \frac{a}{\sqrt{s N}}\sum_{i,\mu\ge2} \xi_i^\mu \mv{z_\mu  k_i} -\frac{1}{ \sqrt{1-s}}\Big( C_1 \sum_i \eta_i \mv{k_i} + C_2 \sum_{\mu\ge 2} \lambda_\mu \mv{z_\mu} \Big)+\\
&+ \frac{a}{\sqrt{s N}}\sum_{i} \xi_i^1 \mv{z_1  k_i} - C_3 \sum_{\mu\ge 2} \mv{z_\mu^2} - C_4 \sum_i\mv{ k_i^2} -C_5 a \sum_i \mv{\xi_i^1 k_i}  \Big].
\ees
We can proceed further by using Wick's Theorem [$\mathbb{E}_{x}xF(x)=\mathbb{E}_{x} (x^2) \cdot \mathbb{E}_{x} \partial_xF(x)$] on the fields $z^1, \,\,\xi^{2\cdots P},\,\,\lambda_\mu,\,\, \eta_i$, obtaining 	
\bes
\frac{d \mathcal{A}(s)}{ds} =&\frac{1}{2N}\qE{\xi,\lambda,\eta}\Big[ \frac{a^2}{N}\sum_{i,\mu\ge 2} \Big(\mv{z_\mu^2  k_i^2}-\mv{z_\mu  k_i}^2\Big) + \frac{a^2}{N} \mv{\big(\sum_i \xi_i^1 k_i\big)^2} -C_1 ^2\sum_i\Big(  \mv{k_i^2}-\mv{k_i}^2\Big) \\
& -C_2^2\sum_{\mu\ge 2}\Big(  \mv{z_\mu^2}-\mv{z_\mu}^2\Big)-C_3\sum_{\mu\ge 2} \mv{z_\mu^2} - C_4 \sum_i\mv{ k_i^2} -C_5 a \sum_i \mv{\xi_i^1 k_i} \Big].
\ees
Using the definition of the order parameters \eqref{eq:orderparam} we can write $d_s \mathcal{A}(s)$ as
\bes
\frac{d \mathcal{A}(s)}{ds} =&\frac{1}{2}\qE{\xi,\lambda,\eta}\Big[a^2\alpha \mv{q_{11}p_{11}} +a^2 \mv{m_1^2}-a^2\alpha\mv{q_{12}p_{12}}-C_1^2\mv{q_{11}}+C_1^2\mv{q_{12}}
+\\
&-C_2^2\alpha\mv{p_{11}}+C_2^2\alpha\mv{p_{12}}-\alpha C_3 \mv{p_{11}}-C_4 \mv{q_{11}} - aC_5\mv{m_1}\Big].
\ees
It is now convenient to fix the free scalars $C_{1,..,5}$ as
\beq
C_1^2=a^2\alpha p, \quad C_2^2=a^2  q, \quad C_3=a^2( W- q),  \quad C_4=a^2\alpha ( X - p),  \quad  C_5=2 m a,
\label{eq:interpcoeff}
\eeq
such that we can recast the streaming $d_s\mathcal{A}(s)$ as
\bes
\frac{d \mathcal{A}(s)}{ds} =&\frac{1}{2}\qE{\xi,\lambda,\eta}\Big[a^2\alpha \mv{(q_{11}- W)(p_{11}- X)} +a^2 \mv{(m_1- m)^2}-a^2\alpha\mv{(q_{12}- q)(p_{12} -  p)}\Big]+\\
&+\frac{\alpha a^2}{2}( q p- W X)-\frac{a^2}{2} m^2.
\label{eq:streamfunc}
\ees
\begin{Remark}\label{Marco6}
When requiring replica symmetry, we have that $\langle q_{11} \rangle \to W$, $\langle p_{11} \rangle \to X$, $\langle m_1 \rangle \to m$, $\langle q_{12} \rangle \to q$ and $\langle p_{12} \rangle \to p$, hence the evaluation of the integral in eq. (\ref{eq:sumrule}) becomes trivial as the r.h.s. of eq. (\ref{eq:streamfunc}) reduces to
\bes\label{famolafinita}
d_s \mathcal{A}(s)=\frac{\alpha a^2}{2}( q p- W X)-\frac{a^2}{2} m^2
\ees
that does not depend on $s$ any longer.
\end{Remark}

\bigskip

\begin{figure}[t!]
	\centering
	\includegraphics[scale=0.55]{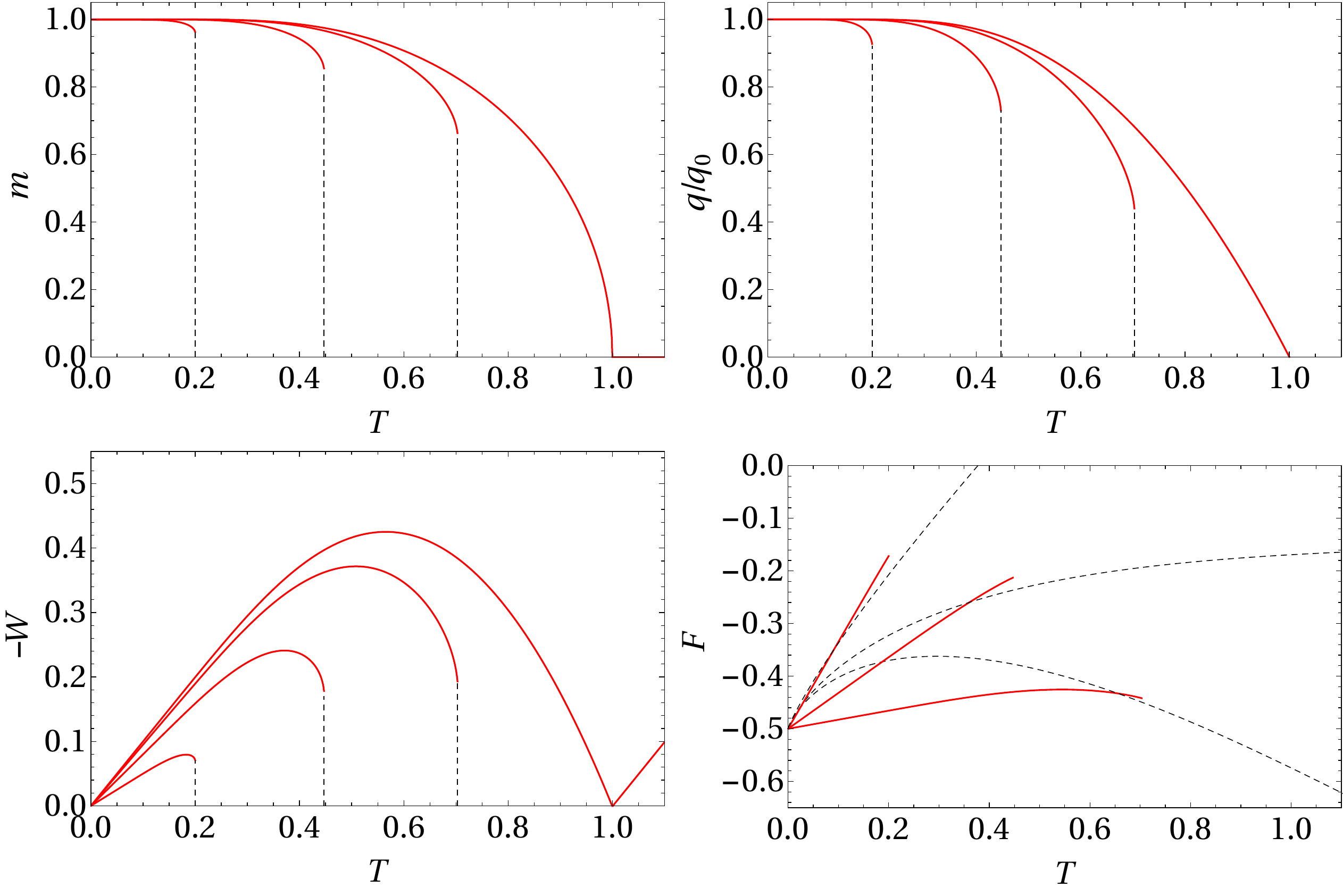}
	\caption{{\bfseries Retrieval state solution for the order parameters and free energy at $t=1000$.} First row: on the left, the plot shows the Mattis magnetization $m$ as a function of the temperature for various storage capacity values ($\alpha=0$, $0.05$, $0.2$ and $0.5$, going from the right to the left). The vertical dotted lines indicates the jump discontinuity identifying the critical temperature $T_c(\alpha)$ which separates the retrieval region from the spin-glass phase; on the right, the plot shows the solutions of the non-diagonal overlap $q$ (normalized to the zero-temperature value $q_0=q(T=0)$), for the same capacity values. The solution is computed in the retrieval region ({\it i.e.} $T<T_c(\alpha)$). Second row: on the left, the plot shows the solution for the diagonal overlap $-W$ in the retrieval region for $\alpha=0$, $0.05$, $0.2$ and $0.5$, finally, on the right the plot shows the free-energy as a function of the temperature for various storage capacity values ($\alpha=0.05$, $0.2$ and $0.5$, going from the bottom to the top) for both the retrieval (red solid lines) and spin-glass (black dashed lines) states.}\label{fig:jumpVST}
\end{figure}

We must now evaluate the one-body contribution $\mathcal{A}(s=0)$: this can be done by directly setting $s=0$ in \eqref{eq:interpfunc}
\bes
\mathcal{A}(s=0)=&\frac{1}{N}\qE{\xi,\eta,\lambda} \ln\sum_{\sigma}\int\dm[z,\phi] \exp \Big[ C_1 \sum_i \eta_i k_i + \frac{C_4}{2} \sum_i k_i^2 +\frac{C_5 a}{2}\sum_i \xi_i^1 k_i  +\\
& + C_2 \sum_{\mu\ge 2} \lambda_\mu z_\mu + \frac{C_3}{2}\sum_{\mu\ge 2} z_\mu^2\Big].
\ees
Performing standard Gaussian integrations we obtain
\bes
\mathcal{A}(s=0)=&-\frac{\alpha}{2}\ln(1-C_3)-\frac{1}{2}\ln(1-C_4 b^2)+\frac{\alpha}{2}\frac{C_2^2}{1-C_3}+\frac{C_4}{2}+\qE{\eta}\ln\cosh\Big[\frac{C_1\eta+\frac{C_5 a}{2}}{1-C_4 b^2}\Big]+\\
&+b^2\frac{C_1^2+C_4^2+\frac{C_5^2a^2}{4}}{1-C_4 b^2}+\ln 2.
\label{eq:onebody}
\ees
Keeping in mind the expressions for the parameters $C_1,...,C_5$ as prescribed in the relations \ref{eq:interpcoeff}, by plugging eq.~(\ref{famolafinita}) and eq.~\eqref{eq:onebody} into the sum rule \eqref{eq:sumrule}
we finally get an expression for the quenched pressure of the model (\ref{new-model}) in terms of the replica-symmetric order parameters
\bes
A_{\RS}(\alpha,\beta,t)=&\frac{\alpha a^2}{2}\big( q p- W X\big)-\frac{a^2}{2}  m^2-\frac{\alpha}{2}\ln\big[1-a^2( W- q)\big]-\frac{1}{2}\ln\big[1-a^2b^2\alpha( X- p)\big]+\\
&+\frac{\alpha}{2}\frac{a^2 q}{1-a^2( W- q)}+\frac{\alpha a^2}{2}\big( X- p\big)+\frac{a^2b^2}{2}\cdot \frac{\alpha  p+ m^2 a^2+a^2\alpha^2( X- p)^2}{1-a^2b^2\alpha( X- p)}+\\
&+\ln 2+\qE{\eta}\ln\cosh\Big[\frac{a\eta\sqrt{\alpha  p} + m a^2}{1-\alpha a^2b^2( X- p)}\Big].
\ees
To match exactly the notation in \cite{Albert2} there is still a short way to go: it is convenient to re-scale $ m$, $ p$ and $ X$ as
\beq
 X\to\frac{\beta^2}{a^2} X,\quad  p\to\frac{\beta^2}{a^2} p,\quad  m\to\frac{\beta}{a^2} m,
\eeq
as this allows us to introduce the composite order parameter $ \Delta=1-\alpha \beta^2 b^2( X- p)$ used in \cite{Albert2}.
\newline
After these transformations, remembering the definition of the free energy (see (\ref{freeEnergy})) and the definition of $(a,b)$ (see \eqref{eq:defspins}), we obtain exactly the same expression for the quenched free energy as that achieved in \cite{Albert2} via the replica trick, as stated by the next main
\begin{Theorem}
In the infinite volume limit, the replica symmetric free energy related to the neural network defined by eq. (\ref{new-model}) can be expressed in terms of the natural order parameters of the theory (see def.s (\ref{eq:orderparam})) as
\bes
F_{\RS}(\alpha,\beta,t)=&-\frac{\beta m ^2}{2(1+t)} \Big(1+\frac{t}{\Delta}\Big)-\frac{(1+t)(\Delta-1)}{2t}\beta W - \frac{\alpha\beta^2}{2} p ( W- q)\\
&-\frac{\alpha}{2}\Big(\log[1-\beta(1+t)( W- q)]+\frac{ q\beta^2(1+t)}{1-\beta(1+t)( W- q)}\Big)-\frac{(1+t)(1-\Delta)\beta}{2 t \Delta}\\
&-\frac{\log \Delta}{2}-\frac{\alpha \beta  p t}{2(1+t)\Delta}+  \qE{\eta} \log \cosh \Big[\frac{\beta }{\Delta}(m+\sqrt{\alpha p}\eta)\Big]+\log 2.
\label{eq:qRSfreeenergy}
\ees
\end{Theorem}
\begin{Proposition}
Using the standard variational principle $\vec{\nabla}F_{\RS}=0$ on the free energy (\ref{eq:qRSfreeenergy}), namely by extremizing the latter over the order parameters, we obtain the following set of self-consistent equations for these parameters, whose behavior is outlined in the plots of Fig.~\ref{fig:jumpVST}.
\bes
	m &=\frac{1+t}{\Delta+t}\qE{\eta} \tanh\Big[\frac{\beta}{\Delta}(m+\sqrt{\alpha p}\eta)\Big],\\
	p &=\frac{q(1+t)^2}{[1-\beta(1+t)(W-q)]^2},\\
	\Delta &=1+\frac{\alpha t}{1-\beta(1+t)(W-q)},\\
	q &=W+\frac{t}{\beta (1+t)\Delta}-\frac{1}{\Delta^2}\qE{\eta}\cosh^{-2}\Big[\frac{\beta}{\Delta}(m+\sqrt{\alpha p}\eta)\Big],\\
	W \Delta^2 &=1-\frac{t\Delta}{\beta(1+t)}+\frac{\alpha p t^2-m^2 t(t+2\Delta)}{(1+t)^2} -\frac{2\alpha\beta p t}{(1+t)\Delta}\qE{\eta}\cosh^{-2}\Big[\frac{\beta}{\Delta}(m+\sqrt{\alpha p}\eta)\Big].
	\label{eq:sceqs}
\ees
\end{Proposition}
\begin{Remark}
We stress that we obtained exactly the same self-consistencies previously appeared in \cite{Albert2}, thus all the consequences stemming by them, as reported in that paper, are here entirely confirmed.
\end{Remark}

\section{Study of the overlap fluctuations}
As proved in the previous section, the reinforcement$\&$removal algorithm makes the retrieval region in the $(\alpha,\beta)$ plane wider and wider as $t$ is increased (see Fig. \ref{fig:criticallines}). As the retrieval region pervades the spin-glass region, one therefore naturally wonders whether the opposite boundary of the spin-glass region (namely the critical line depicting the transition where ergodicity breakdowns) is as well deformed.
%
To address this point, we now study the behavior of the overlap fluctuations, suitably centered around the thermodynamic values of the overlaps and properly rescaled in order to allow them to diverge when the system approaches the critical line. In fact, they are meromorphic functions and their poles  identify the evolution of the critical surface $\beta_c(\alpha,t)$ (if any).
\newline
It is worth recalling that the critical line for the standard Hopfield model \cite{Hopfield} as predicted by the AGS theory \cite{Amit} is $\beta_c (\alpha, t=0) = (1+\sqrt{\alpha})^{-1}$.

\subsection{Guerra's interpolating framework for the overlap fluctuations}

The idea is the same exploited in the previous section, namely to use the generalized Guerra's interpolation scheme (see eq. (\ref{eq:interpfunc})) to evaluate the evolution of the order parameter's correlation functions from $s=0$ (where they do not represent the real fluctuations in the system, but their evaluation should be possible) up to $s=1$ (where they reproduce the true fluctuations). To achieve this goal for the generic correlation function $O$, we need to evaluate the Cauchy condition $\langle O(s=0) \rangle$ and the derivative $\partial_s \langle O(s) \rangle$. However, in contrast with the previous section where we imposed replica symmetry, here -as we just want to infer the critical line- we impose ergodic behavior, namely, we assume that the system is approaching this boundary from the high fast-noise limit. This allows us to set all the mean values of the overlaps to zero and to achieve explicit solutions.
\begin{Definition}
The centered and rescaled overlap fluctuations $\theta_{lm}$ and $\rho_{lm}$ are introduced as
\bes
\theta_{lm}&=\sqrt{N}\big[q_{lm}-\delta_{lm}W-(1-\delta_{lm})q\big]\\
\rho_{lm}&=\sqrt{P}\big[p_{lm}-\delta_{lm}X-(1-\delta_{lm})p\big].
\ees
\end{Definition}
\begin{Remark}
As we will address the problem of the overlap fluctuations in the ergodic region, the signal is absent, thus there is no need to introduce a rescaled Mattis order parameter: only the boundary between the ergodic region and the spin-glass region is under study here.
\end{Remark}
\begin{Proposition}
It is convenient to introduce the $r-$replicated interpolating pressure $\calA_J^r(s)$, where we further added a source field $J$, coupled to an observable $O$ (that is a smooth function of the neurons of the $r$-replicas) as
\bes\label{eq:repinterp}
\calA^r_J(s)=&\qE{\xi,\eta,\lambda} \ln\sum_{\sigma_R}\int\dm[z_R,\phi_R] \exp \Big[ \sqrt{s}\frac{a}{\sqrt{N}}\sum_{l=1}^r\sum_{i,\mu} z_\mu^{(l)} \xi_i^\mu k_i^{(l)}+J\hat{O}\\
& + \sqrt{1-s}\Big( C_1 \sum_{l=1}^r\sum_i \eta_i k_i^{(l)} + C_2 \sum_{l=1}^r\sum_{\mu} \lambda_\mu z_\mu^{(l)} \Big)+\frac{1-s}{2}\Big( C_3 \sum_{l=1}^r\sum_{\mu} (z_\mu^{(l)})^2 + C_4 \sum_{l=1}^r\sum_i (k_i^{(l)})^2\Big)\Big].
\ees
where $k_i$ is the same as in Definition $5$ and the interpolation constants $C_{1,2,3,4}$ are the same given in the previous section (see eq. (\eqref{eq:interpcoeff})).
\end{Proposition}
By definition
\beq
\mv{O(s)}=\evalat{\der{\calA^r_J(s)}{J}}{J=0}, \quad \quad \partial_s\mv{O(s)}=\evalat{\der{(\partial_s \calA^r_J)}{J}}{J=0}.
\label{eq:repgenfunc}
\eeq
Therefore, in order to evaluate the fluctuations of $O$ we need to evaluate first $\partial_s \calA^r_J$ and, by a routine calculation, we get
\beq
\partial_s \calA^r_J=\frac{1}{2}\sqrt{\alpha}\beta(1+t)\sum_{l,m=1}^r\Big[\mv{g_{l,m}}-\mv{g_{l,m+r}}\Big],\quad \quad g_{l,m}=\theta_{l,m}\rho_{l,m}.
\eeq
To evaluate the fluctuations of a general operator $O$, function of $r-$replicas, we must use the results \eqref{eq:repgenfunc} and perform the same rescaling that we did in the previous section, namely
\beq
(X,p)\to\frac{\beta^2}{a^2}(X,p).
\eeq
Overall this brings to the next
\begin{Proposition} Given $O$ as a smooth function of $r$ replica overlaps $\left( q _ { 1 } , \ldots , q _ { r } \right)$ and $\left( p _ { 1 } , \ldots , p _ { r } \right) ,$ the following streaming equation holds:
\beq
d_{\tau}  \mv{O}=\frac{1}{2}\sum _ { a , b } ^ { r } \mv {O \cdot g_ { a , b }}- r \sum _ { a = 1 } ^ { r } \mv{O \cdot g _ { a , r + 1 } } +  \frac{r ( r+ 1 )}{2} \mv{ O \cdot g_ { r + 1 , r + 2 } } -\frac{r}{2} \mv{ O \cdot g_ { r + 1 , r + 1 } },
\label{eq:fluctstream}
\eeq
where we used the operator $d_{\tau}$ defined as
\beq
d_{\tau} = \frac{1}{\beta (1+t)\sqrt{\alpha}}\frac{d}{ds},
\eeq
in order to simplify calculations and presentation.
\end{Proposition}

\subsection{Criticality and ergodicity breaking}

To study the overlap fluctuations we must consider the following correlation functions (it is useful to introduce and link them to
capital letters in order to simplify their visualization):
\bes
\mv{ \theta_{12}^2 }_s &= A(s), &\mv{\theta_{12}\theta_{13}}_s &= B(s), &\mv{ \theta_{12}\theta_{34}}_s &= C(s), \\
\mv{ \theta_{12}\rho_{12} }_s &= D(s), &\mv{\theta_{12}\rho_{13} }_s &= E(s), &\mv{\theta_{12}\rho_{34}}_s &= F(s), \\
\mv{ \rho_{12}^2}_s &= G(s), &\mv{\rho_{12}\rho_{13} }_s &= H(s), &\mv{\rho_{12}\rho_{34}}_s &= I(s),\\
\mv{ \theta_{11}^2 }_s &= J(s), &\mv{\theta_{11}\rho_{11} }_s &= K(s), &\mv{\rho_{11}^2}_s &= L(s),\\
\mv{ \theta_{11}\theta_{12} }_s &= M(s), &\mv{\theta_{11}\rho_{12} }_s &= N(s), &\mv{\rho_{11}\theta_{12}}_s &= O(s),\\
\mv{\rho_{11}\rho_{12}}_s &= P(s),&\mv{\theta_{11}\rho_{22} }_s &= Q(s),&\mv{\theta_{11}\theta_{22} }_s &= R(s).\\
\mv{\rho_{11}\rho_{22}}_s &= S(s),
\ees
Since we are interested in finding the critical line for ergodicity breaking {\em from above} we can treat $\theta_{a,b},\rho_{a,b}$ as Gaussian variables with zero mean (this allows us to apply Wick-Isserlis theorem inside averages) as we can also treat both the $k_i$ and $z_\mu$ as zero mean random variables in the ergodic region (thus all averages involving uncoupled fields are vanishing): this considerably simplifies the evaluation of the critical line (as expected since we are approaching criticality from the {\em trivial} ergodic region \cite{BarraGuerra-JMP2008}).
\newline
We can thus reduce the analysis to
\bes
\mv{ \theta_{12}^2 }_s &= A(s), &\mv{ \theta_{12}\rho_{12} }_s &= D(s), &\mv{ \rho_{12}^2 }_s &= G(s),\\
\mv{ \theta_{11}^2 }_s &= J(s), &\mv{\theta_{11}\rho_{11} }_s &= K(s), &\mv{\rho_{11}^2}_s &= L(s),\\
\mv{\theta_{11}\rho_{22} }_s &= Q(s),&\mv{\theta_{11}\theta_{22} }_s &= R(s), &\mv{\rho_{11}\rho_{22}}_s &= S(s).
\ees
According to \eqref{eq:fluctstream} and to the previous reasoning we obtain:
\bes
d_\tau {A}&=2AD, \\
d_\tau {D}&=D^2+AG,\\
d_\tau {G}&=2GD.
\label{eq:pde}
\ees
Suitably combining $A$ and $G$ in \eqref{eq:pde} we can write
\beq
d_\tau \ln \frac{A}{G}=0\implies A(\tau)=r^2G(\tau), \quad r^2=\frac{A(0)}{G(0)}.
\eeq
Now we are left with
\bes
d_\tau{D}&=D^2+r^2G^2,\\
d_\tau{G}&=2GD.
\label{eq:pde2}
\ees
The trick here is to complete the square by summing $d_\tau D + r d_\tau G$ thus obtaining
\bes
d_\tau Y&=Y^2,\\
Y&=D+rG,\\
d_\tau{G}&=2G(Y-rG).
\label{eq:pde3}
\ees
The solution is trivial and it is given by
\beq
Y(\tau)=\frac{Y_0}{1-\tau Y_0},\quad Y_0=D(0)+\sqrt{A(0)G(0)}.
\label{eq:fluctY}
\eeq
\begin{figure}[h!]
	\centering
	\includegraphics[scale=0.5]{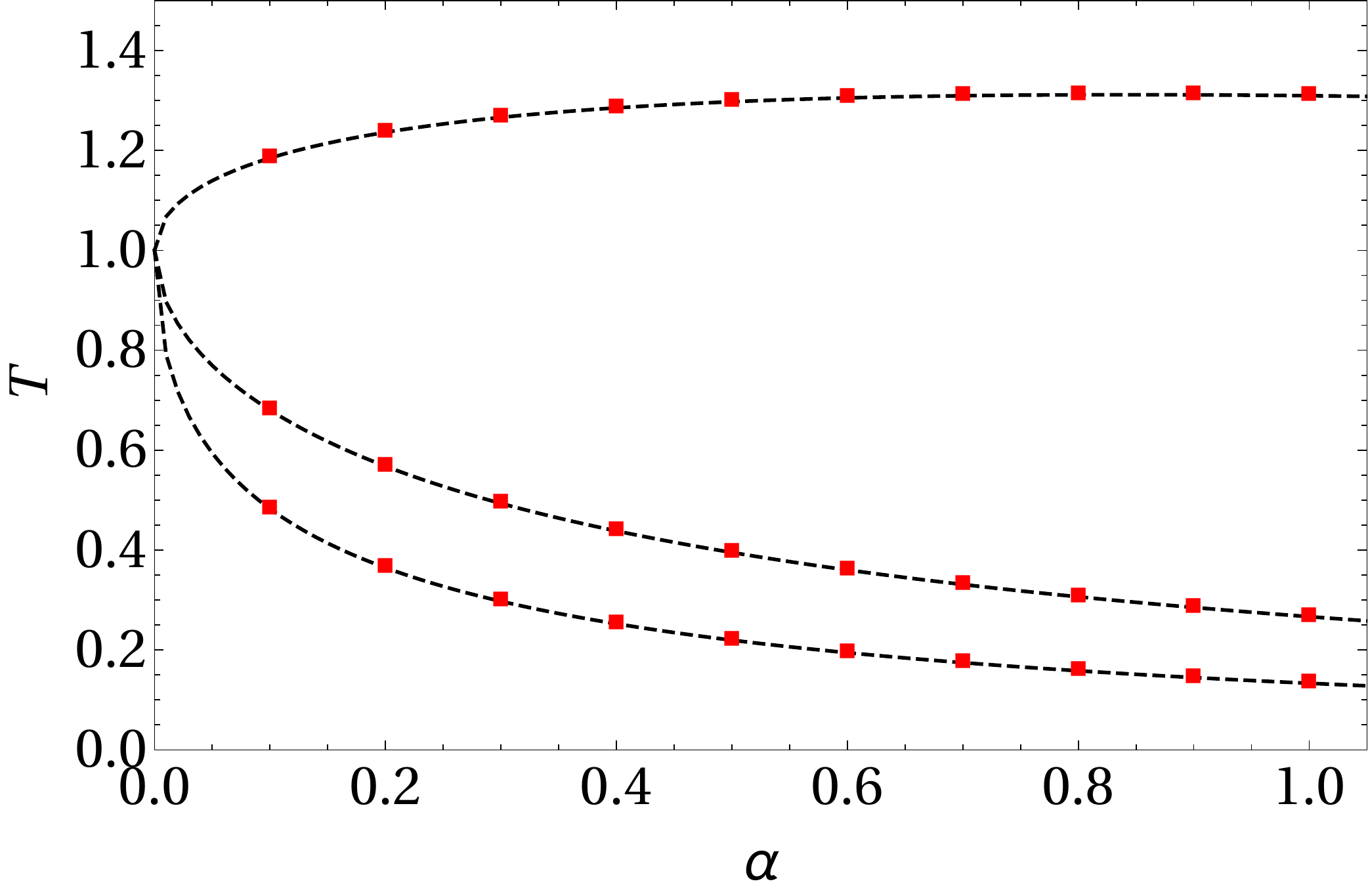}
	\caption{{\bfseries Ergodicity breaking critical line.} The plot shows a comparison between the theoretical predictions (black dashed lines) for the ergodicity breaking critical line according to Eq. \eqref{eq:ergodicityline} and numerical solutions for spin glass states (red markers). The latter are evaluated by solving the self-consistency equations with $m=0$ with  $\alpha$ fixed and searching for the temperature $T$ above which the solution has $q=0$. Going from top to bottom of the plot, the sleep extent is $t=0.1$, $1$ and $2$.}\label{fig:ergo}
\end{figure}
So we are left with the evaluation of the correlations at $s=0$: namely the Cauchy conditions related to the solution coded in eq. (\ref{eq:fluctY}).
To this task we introduce a one-body generating function for the momenta of $z,k$: this can be done by setting inside \eqref{eq:repinterp} $s=0,r=1$ and adding source fields $(j_i, J_{\mu})$ coupled respectively to $(k_i,z_\mu)$, with $i \in (1,...,N),\,\mu \in (1,...,P)$. Since we are approaching the critical line from the high fast noise limit we can set $m,p,q=0$ (when we explicitly make use of the coefficients \eqref{eq:interpcoeff}), overall writing
\bes\label{eq:onebodygenf}
F(j,J)=&\ln\sum_{\sigma}\int\dm[z,\phi] \exp \Big[ \sum_i j_i k_i + \sum_{\mu} J_\mu z_\mu +\frac{a^2 W}{2}\sum_{\mu} z_\mu^2 + \frac{1-\Delta}{2b^2} \sum_i k_i^2\Big].
\ees
Clearly, we took great advantage in approaching the ergodic region from above, since even the one-body problem (for the Cauchy condition) has been drastically simplified: showing only the relevant terms in $j,J$ we have
\bes
F(j,J)=\frac{b^2\Delta +1}{2\Delta^2}\sum_i j_i^2+\frac{1}{2(1-a^2W)}\sum_\mu J_\mu^2 +O(j^3).
\ees
As anticipated, all the observable averages needed at $s=0$ can now be calculated simply as derivatives of $F(j,J)$, thus the $s=0$ correlation functions are finally given by
\bes
D(0)&=\sqrt{NP}\evalat{\big(\partial_j F\big)^2\big(\partial_J F\big)^2}{j,J=0}=0,\\
A(0)&=\evalat{\big(\partial^2_j F\big)^2}{j,J=0}=\Big[\frac{\beta(1+t)-t\Delta}{\beta(1+t)\Delta^2}\Big]^2=W^2,\\
G(0)&=\evalat{\big(\partial^2_J F\big)^2}{j,J=0}=(1-\beta(1+t)W)^{-2}.
\ees
Inserting this result in \eqref{eq:fluctY}, we get
\bes
Y(\tau)=\frac{W}{1-\beta(1+t)W-\tau W}.
\ees
Upon evaluating $Y(\tau)$ for $\tau=\beta(1+t)\sqrt{\alpha} s,\, s=1$ and reporting the relevant ergodic self-consistent equations we obtain the following system:
\bes\label{polo}
Y(s=1)&=\frac{W}{1-\beta(1+t)W(1+\sqrt{\alpha})},\\
W\Delta^2 &= 1-\frac{t\Delta}{\beta(1+t)},\\
\Delta &=1+\frac{\alpha t}{1-\beta(1+t)W}.
\ees
Since we are interested in obtaining the critical temperature for ergodicity breaking, where fluctuations (in this case $Y$) grow arbitrarily large we can check where the denominator at the r.h.s. of the first eq. (\ref{polo})  becomes zero and recast this observation as follows
\begin{Theorem}
The ergodic region of the model defined by the cost function (\ref{new-model}) is delimited by the following critical surface in the $(\alpha,\beta,t)$ space of the tunable parameters
\begin{equation}\label{eq:ergodicityline}
\beta_c=\frac{1}{1+t}\Big[\frac{\Delta^2}{1+\sqrt{\alpha}}+t\Delta\Big]\quad \text{with}\quad 
\Delta=1+\sqrt{\alpha}(1+\sqrt{\alpha})t.
\end{equation}
\end{Theorem}

\begin{Remark}
At $t=0$, where the model reduces to Hopfield's scenario, the critical surface correctly collapses over the Amit-Gutfreund-Sompolinsky critical line $\beta_c=(1+\sqrt{\alpha})^{-1}$, but in the large $t$ limit the ergodic region collapses to the axis $T=0$: this may have a profound implication, namely that the ergodic region -during the sleep state- {\em phagocytes} the spin-glass region.
\newline
Since we have already seen that also the retrieval region {\em phagocytes} the spin-glass region \footnote{Note that the ergodic line does not affect the retrieval region, they simply {\em fade} one into the other. This is because the critical surface is calculated assuming an ergodic regime (hence, it does not takes into account the signal) and, more importantly, the retrieval region is delimited by a first order phase transition, that is not detected by a second order inspection as that needed for criticality.} this means that spurious states are entirely suppressed with a proper rest, allowing the network to achieve perfect retrieval, as suggested in the pioneering  study by Kanter and Sompolinsky  \cite{KanterSompo}.
\end{Remark}


\section{Conclusions and outlooks}

In recent years Artificial Intelligence, mainly due  to the impressive skills of Deep Learning machines and the GPU-related revolution  \cite{DL1}, has attracted the attention of the whole Scientific Community. In particular, the latter includes mathematicians involved in the statistical mechanics of complex systems which has proved to be a fruitful tool in the investigation of neural networks and machine learning, since the early days (not by chance {\em Boltzmann machines} are named after {\em Boltzmann} \cite{BM1}).
\newline
Among the various fields of Artificial Intelligence where, in the present years, statistical mechanics extensively contributed to the cause (e.g. statistical inference and signal processing \cite{Simona,Lenka1}, combinatorial and computational complexity \cite{Lenka2,Zecchina,Monasson}, supervised or unsupervised learning \cite{Zecchina2,Huang}, deep learning \cite{Chiara,Metha}, compositional capabilities \cite{Agliari-PRL1,Monasson2}, and really much more...) the one we deepened in this work deals with the phenomenon of {\em dreaming and sleeping}\footnote{We point out that dreaming has been recently connected to compositional capabilities \cite{Guardian}, the latter being natural properties of diluted retricted Boltzmann machines \cite{Agliari-Dantoni,Agliari-Immune,Monasson2}.}.
\newline
In the current work we mathematically described the phenomena of reinforcement and remotion, as pioneered by Crick $\&$ Mitchinson \cite{Crick}, by Hopfield \cite{HopfieldUnlearning} and by many others in the neuroscience literature, see e.g \cite{Neuro-1,Neuro0,Neuro1,Neuro2}): interestingly, such mechanisms have been evidenced to lead to an improvement of the retrieval capacity of the system.
In particular, in \cite{Albert2}, we showed that the system reaches the expected upper critical capacity $\alpha_c=1$, still preserving robustness with respect to fast noise. However, the statistical mechanical analysis, set at the standard replica symmetric level of description, was carried out via non-rigorous approaches (e.g., replica trick and numerical simulations).
\newline
In this work we extended a Guerra's interpolation scheme \cite{Barra-JSP2010}, originally developed to deal with the standard Hopfield model (i.e. equipped with the canonical Hebbian synaptic coupling), to deal with this generalization: at first we showed the equivalence of this model with a three-layer spin-glass where some links among different layers are cloned (hence introducing correlation in the network and in the random fields required for the interpolation) and the third, and novel (w.r.t. the standard equivalence between Hopfield models and two-layers Boltzmann machines \cite{BarraEquivalenceRBMeAHN,Barra-RBMsPriors2}), layer is equipped with imaginary real-valued neurons (best suitable to perform spectral analysis\footnote{We plan to report soon on the learning algorithms for this generalized restricted Boltzmann machine, where the properties of the spectral layers will spontaneously shine.}). As a consequence, the resulting interpolating architecture is rather tricky, by far richer than its classical limit yet it turns out to be managable and actually a sum rule for the quenched free energy related to the model can be written and even integrated, under the assumption of replica symmetry: such an expression, as well as those stemming from its extremization for the order parameters, sharply coincides with previous results \cite{Albert2}, confirming them in each detail.
\newline
We remark that such theorems state also the validity of other previous investigation -all replica trick derived- on unlearning in neural networks (see e.g.  \cite{Dotsenko1,unlearning1,KanterSompo}).
\newline
Beyond confirming previous results, we further systematically developed a fluctuation analysis of the overlap correlation functions, searching for critical behaviour, in order to inspect where ergodicity breaks down and in this investigation we found a very interesting result: as long as the Hopfield model is awake, the critical line is the one predicted by Amit-Gutfreund-Sompolinksy (as it should and as it is known by decades). However, as the network sleeps, the ergodic region starts to invade the spin glass region, ultimately destroying the spin glass states entirely, thus allowing the network (at the end of an entire sleep session) to live {\em solely} within a -quite large- retrieval region, surrounded by ergodicity: noticing that at this final stage of sleeping the network approached the Kanter-Sompolinsky model \cite{KanterSompo}, it shines why these Authors called their model {\em associative recall of memory without errors}.

\section*{Acknowledgments}
The Authors acknowledge partial financial fundings by MIUR, via {\em FFABR2018-(Barra)} and via {\em Rete Match - Progetto Pythagoras}  (CUP:J48C17000250006) and by INFN.

\end{document}